\documentclass[journal=jacsat,manuscript=article]{achemso} 
\usepackage[version=3]{mhchem} % Formula subscripts using \ce{}
\usepackage{lineno}[left]
\usepackage[usenames,dvipsnames]{xcolor}
\usepackage{array}
\usepackage{tabularx}
\usepackage{amsmath}
\author{Sohag Biswas}
\author{Bryan M. Wong}
\email{bryan.wong@ucr.edu, Website: http://www.bmwong-group.com}
\affiliation[Unknown University]
{Department of Chemical \& Environmental Engineering, Materials Science \& Engineering Program, Department of Physics \& Astronomy, and Department of Chemistry, University of California-Riverside, Riverside, California 92521, USA}

\title[An \textsf{achemso} demo]
  {High-Temperature Decomposition of Diisopropyl Methylphosphonate (DIMP) on Alumina: Mechanistic Predictions from Ab Initio Molecular Dynamics}

\begin{document}

\begin{abstract}
The enhanced degradation of organophosphorous-based chemical warfare agents (CWAs) on metal-oxide surfaces holds immense promise for neutralization efforts; however, the underlying mechanisms in this process remain poorly understood. We utilize large-scale quantum calculations for the first time to probe the high-temperature degradation of diisopropyl methylphosphonate (DIMP), a nerve agent simulant. Our Born-Oppenheimer molecular dynamics (BOMD) calculations show that the $\gamma$-Al$_2$O$_3$ surface shows immense promise for quickly adsorbing and destroying CWAs. We find that the alumina surface quickly adsorbs DIMP at all temperatures, and subsequent decomposition of DIMP proceeds via a propene elimination. Our BOMD calculations are complemented with metadynamics simulations to produce free energy paths, which show that the activation barrier decreases with temperature and DIMP readily decomposes on $\gamma$-Al$_2$O$_3$. Our first-principle BOMD and metadynamics simulations provide crucial diagnostics for sarin decomposition models and mechanistic information for examining CWA decomposition reactions on other candidate metal oxide surfaces.
\end{abstract}

\section{Introduction}

The neutralization of chemical warfare agents (CWAs) continues to be a pressing area of interest for the safe and effective removal of these hazardous compounds. Among the various CWAs, the most nefarious are organophosphate nerve agents (such as sarin), which contain P=O, P--O--C, and P--C bonds that enable lethal phosphorylating mechanisms.\cite{kim-2011} 
%Exposure to nerve agents at low concentrations, death can happen within one to ten minutes after direct inhalation.
Over the past few decades, a variety of destruction and neutralization methods have been used to safely eliminate these hazardous compounds. 
For example, destruction-based methods (typically pyrolysis) allow a one-step approach for the complete disposal of CWAs at the expense of using specialized equipment under extreme conditions\cite{kim-2011,picard-2019}. On the other hand, neutralization methods offer potentially reversible chemical treatments, leading to possible CWA precursors under less severe conditions.\cite{kim-2011}

Recent studies have shown that metal oxides can effectively destroy CWAs due to their high surface area, a large number of highly reactive edges, corner defect sites, unusual lattice planes, and high surface-to-volume ratios. In particular, metal oxides, such as CaO,\cite{michalkova-2007,wagner-2000} MgO,\cite{michalkova-2004,wagner-1999} ZnO,\cite{mahato-2009,prasad-2007,prasad-2011} TiO$_2$,\cite{hirakawa-2010,prasad-2008,prasad-2009,rama-2012,sato-2011,STENGL-2012,wagner-2008} Al$_2$O$_3$,\cite{kuiper-1976, saxena-2009,saxena-2010,wagner-2001} Fe$_3$O$_4$,\cite{walenta-2020} and CuO\cite{troto-2017} are candidates as adsorbents for enhancing the decomposition of CWAs. 
%In the present work, we focus on the $\gamma$-Al$_2$O$_3$ as an adsorbent, which has a high degree of surface site heterogeneity.\cite{calle-2015,dinge-2004}
In $\gamma$-Al$_2$O$_3$,  Al atoms in the bulk exhibit either a tetrahedral or octahedral coordination. However, depending on the exposed crystallographic surface, Al atoms on the surface can display penta-, tetra, and tri-coordination and exhibit Lewis acidity.\cite{dinge-2004,gu-2018,roy-2012,christ-2013} Because of its  high degree of surface heterogeneity, $\gamma$-Al$_2$O$_3$ offers a high catalytic activity and is a promising candidate for the decomposition of various CWAs.\cite{bermudez-2007} 
%has been used to convert alkanes to olefins,\cite{dixit-2018} dehydrate alcohols,\cite{roy-2012,christ-2013} and chemically activate methane.\cite{chole-2018} As such, alumina has also been suggested as 

%Nerve agents are recognized to be the most nefarious and hazardous among all of the CWAs. These nerve agents are organophosphate compounds comprised of P=O, P=S, P--O--C, or P--C bonds that enable a phosphorylating mechanism.\cite{kim-2011} 
%Exposure to nerve agents at low concentrations, death can happen within one to ten minutes after direct inhalation.
%Due to the hazardous nature of CWAs, simulants of nerve agents, such as dimethyl methylphosphonate (DMMP), diisopropyl methylphosphonate (DIMP), diethyl methylphosphonate (DEMP), trimethyl phosphate (TMP), are considered in  laboratory experiments for mimicking CWAs.
In this work, we present the first \textit{ab initio} molecular dynamics study for probing high-temperature decomposition dynamics of diisopropyl methylphosphonate (DIMP)  on the $\gamma$-Al$_2$O$_3$ surface. 
%we utilize large-scale \textit{ab initio} molecular dynamics simulations to probe  decomposition dynamics of diisopropyl methylphosphonate (DIMP)  on the $\gamma$-Al$_2$O$_3$ surface.
Due to its structural similarity with sarin, DIMP has been used in experiments to mimic the decomposition reaction mechanism of CWAs. DIMP is the only surrogate with the same isopropyl (--O--C$_3$H$_7$) group found in sarin gas (the only structural difference between sarin and DIMP is the substitution of a fluorine for an isopropyl group in the former). Several techniques have also been used to probe DIMP decomposition, including microwave discharge approaches,\cite{bailin-1975} pyrolysis on porous substrates,\cite{gibson-2018} laser heating,\cite{thompson-2019} and thermal decomposition at 700 -- 800 K.\cite{senyurt-2021}
Despite these experimental studies, theoretical analyses of DIMP decomposition on metal oxides are scarce, except for a previous study on mechanism and rates of thermal decomposition of DIMP.\cite{glaude-2002}
%In this work, we present DIMP decomposition on the $\gamma$-Al$_2$O$_3$ surface because of its high degree of surface heterogeneity and high catalytic activity.\cite{calle-2015,dinge-2004,dinge-2002} 
In this work, we utilize large-scale \textit{ab initio} molecular dynamics simulations to probe the adsorption dynamics and time scales of DIMP decomposition at various temperatures. In addition, we present new \textit{ab initio}-based metadynamics simulations to calculate free-energy barriers for various bond breaking decomposition reactions of DIMP. These computational techniques allow us to (1) predict activation energies and detailed mechanistic pathways at various temperatures and (2) establish accurate sarin decomposition models on metal-oxide surfaces to guide experimental efforts for neutralizing DIMP.

\section{Simulation Details}
\subsection{Molecular Dynamics Simulations}
Density functional theory (DFT) based Born-Oppenheimer molecular dynamics (BOMD) simulations were carried out using the CP2K\cite{joost_cpc_2005} software suite. We have specifically chosen to use this software package since the implementation of linear-scaling Kohn-Sham approaches in CP2K allows robust and efficient electronic structure calculations for large systems. 
The Perdew-Burke-Ernzerhof (PBE)\cite{perdew-1996} was used for the DFT calculations with Grimmes's D3 method to account for dispersion forces.\cite{grimme-2006}
To obtain reasonable accuracy, we utilized a DZVP (double zeta valence polarized) basis set for Al in the DFT calculations and the TZV2P basis set for C, O, H, and P atoms with the Goedecker, Teter, and Hutter (GTH) pseudopotentials\cite{goedcker_prb_1996,goedcker_prb_1998} for atomic core electrons.
A similar basis set for Al atoms was also used to accurately calculate methane activation\cite{chole-2018} and alkane dehydrogenation\cite{dixit-2018} on the $\gamma$-Al$_2$O$_3$ surface.
The orbital transformation method with an electronic gradient tolerance value of $1 \times 10^{-5}$ atomic units was adopted as the convergence criteria for the SCF cycle.\cite{joost_jcp_2003}
A kinetic energy cut-off of 400 Ry for the auxiliary plane-wave basis with a 0.5 fs timestep was employed to integrate the equations of motion. 
The initial guess was furnished by the stable predictor-corrector extrapolation method at each molecular dynamics step.\cite{kolafa_jcc_2004,kuhne_prl_2007} 
We have carried out several molecular dynamics simulations in the 200 -- 1000$^{\circ}$ C temperature range in increments of 100$^{\circ}$ C (i.e., a total of nine independent trajectories were considered). 
\begin{figure}
    \centering
    \includegraphics[width=13cm]{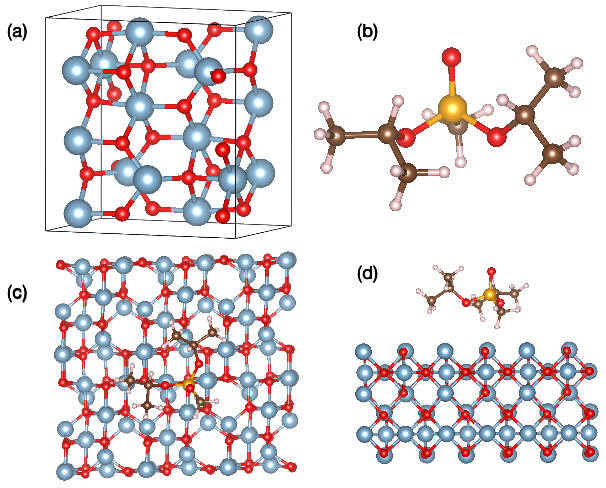}
    \caption{(a) Bulk structure of $\gamma$-Al$_2$O$_3$. (b) Molecular structure of diisopropyl-methylphosphonate (DIMP). (c) Top view of DIMP adsorbed on the  $\gamma$-Al$_2$O$_3$ surface. (d) Side view of  DIMP adsorbed on the $\gamma$-Al$_2$O$_3$ surface. The blue, red, orange, brown, and light pink spheres represent Al, O, P, C, and H atoms, respectively. }
    \label{fig1}
\end{figure}
All simulations were carried out using a Nose-Hoover chain thermostat with the canonical ensemble (NVT).\cite{Nose_jcp_1994,Hoover_pra_1985}

The nonspinel model of the bulk $\gamma$-alumina unit cell (shown in Figure \ref{fig1}a) was built using the crystallographic model by Dinge et al.\cite{dinge-2004,dinge-2002}, which has been shown to match well with experimental structural parameters.\cite{wilson1979}
%Specifically, the crystallographic lattice parameters of $\gamma$-alumina are $a=5.587$ {\AA}, $b=8.413$ {\AA}, and $c=8.068$ {\AA}, with a P21/m space group.\cite{krokidis-2001
Our calculated cell parameters ($a=7.90$ {\AA}, $b=7.93$ {\AA}, and $c=8.07$ {\AA})\cite{dinge-2002,dinge-2004} for bulk $\gamma$-alumina are in excellent agreement and within 2$\%$ of the experimental cell parameters ($a=b=7.96$ {\AA} and $c=7.81$ {\AA}).\cite{wilson1979}
%The structural parameters of this model are close to the experimental ones (within 2$\%$ of our GGA-based DFT calculations).}
%, but calculated bulk electronic properties differ from experimental results.}
%Therefore, this model can not replicate the complexity of $\gamma$-alumina completely, it represents a good compromise between structure reliability and the sizes feasible with quantum chemical calculations. 
In our simulations, we used the (100) facet of this model, which is the lowest-energy facet reported for this material.\cite{dinge-2004} It is worth mentioning that our calculations represent a simplified model of the metal-oxide surface, and impurities (such as H$_2$O, OH, SOx, etc.) may be present and play a crucial role in its reactivity. Nevertheless, our large-scale BOMD calculations still provide critical atomistic insight into the reactivity of the original pristine material, which serves as a baseline for comparing its catalytic activity against other mixed metal oxide surfaces (and their associated impurities), which we save for future studies.

For our NVT simulations, a $3 \times 1 \times 2$ supercell of $\gamma$-alumina and a single DIMP molecule (Figure \ref{fig1}b) containing a total of 268 atoms were used. 
We introduced a single DIMP molecule 5 -- 6 {\AA} above the center of the alumina slab along the $y$-direction, as shown in Figure \ref{fig1}c. Periodic boundary conditions were applied in the $x$ and $z$ directions. Thus, the $xz$ plane of the slab is parallel to the surface, and the $y$-axis forms the surface normal where DIMP interacts with the alumina surface. 
We introduced a vacuum layer of 15 {\AA} on top of the surface to avoid any significant overlap between the electronic density of periodically translated cells.
Our BOMD simulations show that the DIMP molecule moves extremely fast and explores a large space of configurations. To limit this exploration to regions where DIMP dissociation on alumina might occur, it is essential to restrict the movement of the DIMP molecule. 
Specifically, an external spherical potential, which only acts on the DIMP molecule, was placed at the center of the system. Due to the computationally expensive nature of these simulations, we performed 12 ps NVT simulations for each trajectory.
We also calculated adsorption energies using conventional geometry optimizations using the Broyden-Fletcher-Gold-farb-Shanno (BFGS) minimization algorithm until the forces converged to $4.0 \times 10^{-4}$ Bohr with an SCF convergence criteria of $1 \times 10^{-5}$ au. For the geometry optimization, we used a relatively small supercell ($1 \times 2 \times 2$) compared to our BOMD simulations. 

\subsection{Metadynamics Simulations}
\begin{figure}
    \centering
    \includegraphics[width=8cm]{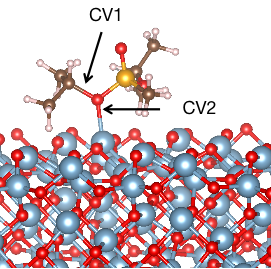}
    \caption{Collective variables (CVs) used to describe the adsorption dynamics of DIMP on the alumina surface.}
    \label{cv}
\end{figure}

Throughout our BOMD simulations, we did not observe any decomposition of DIMP on alumina within the 200 -- 600$^{\circ}$ C temperature range (decomposition did occur at higher temperatures, which is discussed later in this paper). 
At lower temperatures, the high computational cost of these simulations only permits explorations of short time intervals; therefore, BOMD simulations alone do not permit a routine exploration of the complete reaction dynamics. One can sidestep this limitation by using advanced computational techniques such as metadynamics simulations (MetaD) \cite{D0CP03832F}, which we describe further below. In short, MetaD bypasses the sampling limitations of traditional molecular dynamics by applying a history-dependent biasing potential as a function of time to enable efficient sampling of the free energy surface. The free energy surface itself is defined over a set of collective variables (CVs), which are carefully chosen to provide a complete description of the system's slow degrees of freedom. The decomposition process is characterized by the breaking of various bonds within the DIMP molecule on the alumina surface. For this reason, we utilized two CVs, shown in Figure \ref{cv}, for our accelerated sampling simulations.
The CV specifies the coordination number (CN) during the MetaD simulations, which is expressed as a function of the distance between two atoms:
\begin{equation}
\label{cn1}
    \text{CV} \text{  or  } \text{CN} = {{1-\left({{d_{\text{AB}}} \over {d_0}}\right)^p}\over{1-\left({{d_{\text{AB}}}\over{d_0}}\right)^{p+q}}},
\end{equation}
where $d_\text{AB}$ is the distance between atoms A and B, and $d_0$ is the reference distance or fixed cut-off parameter (this parameter characterizes the standard bond distance between atoms A and B). The variables $p$ and $q$ in Equation \ref{cn1} are constants, which were fixed to $p=q=6$.

In this work, we chose CV1 to be the distance between the C and O atoms within DIMP, and CV2 as the distance between the O atom in DIMP and an Al atom on the alumina surface, as shown in Figure \ref{cv}.  MetaD simulations were carried out by depositing Gaussians with heights of 0.02 and 0.001 Hartree for the 200 -- 500$^{\circ}$ C and 600 -- 1000$^{\circ}$ C temperature ranges, respectively. The widths of the Gaussians were set to 0.1 for both the CV1 and CV2 simulations. In this work, well-tempered MetaD (wt-MetaD) simulations were used \cite{barducci-2008,BISWAS2021115624} with the deposition rate of the Gaussian hills set to 20 steps. The well-tempering was implemented  using a Gaussian height damping factor of $\Delta$T such that the ratio $\frac{\Delta \text{T} + \text{T}}{\text{T}} $ was equal to 6.  We performed two sets of wt-MetaD simulations for each temperature from different initial conditions, which were extracted from the pre-equilibrated BOMD simulations. We confirmed a representative sampling and convergence of the reactant, product, transition state, and free energy differences, particularly for the dissociation of the C--O bond in DIMP on the alumina surface. 

\section{Results and Discussions}
\subsection{Adsorption of DIMP on the Alumina Surface}
\begin{figure}
    \centering
    \includegraphics[width=15cm]{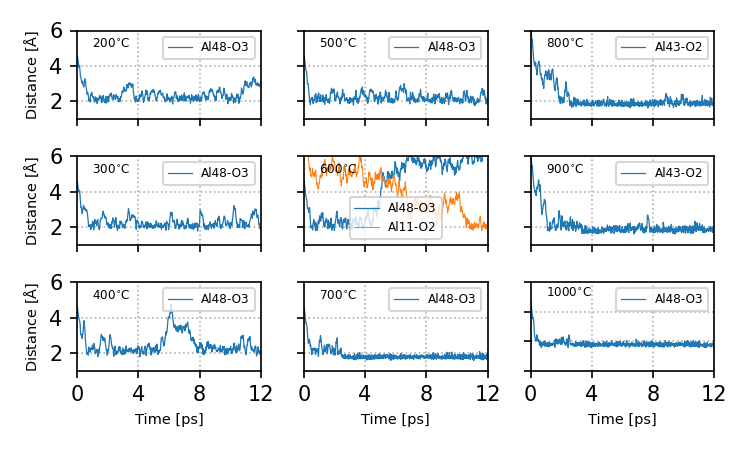}
    \caption{Time-dependent fluctuations of various Al--O distances, which confirm DIMP adsorption throughout the BOMD simulations for all temperatures.}
    \label{dist1}
\end{figure}
\begin{figure}
    \centering
    \includegraphics[width=15cm]{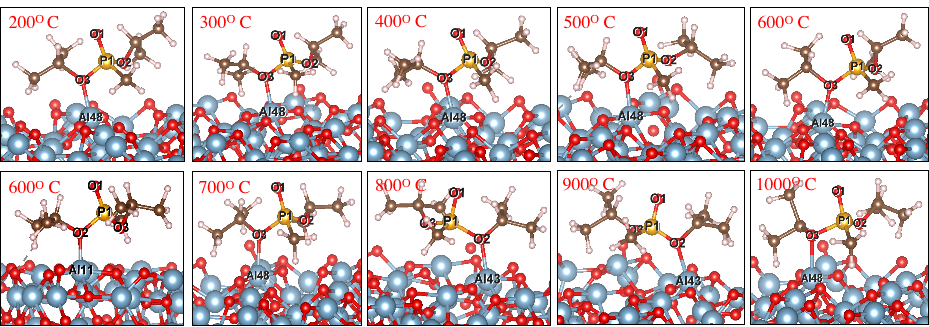}
    \caption{Snapshots illustrating the adsorption of DIMP on the alumina surface. From 200 -- 500$^{\circ}$ C, adsorption occurs via the formation of the Al48--O3 bond. At 600$^{\circ}$ C, adsorption proceeds via the formation of the Al48--O3 and Al11--O2 bonds. At 700 and 1000$^{\circ}$ C, the Al48--O3 bond is formed during the adsorption process, and from 800 -- 900$^{\circ}$ C, an Al43--O2 bond is formed.}
    \label{ads}
\end{figure}

Metal oxide surfaces are typically catalytic and can adsorb and subsequently decompose CWAs into benign products.
Prior experiments\cite{mitchell-1997} and theoretical\cite{bermudez-2007} calculations have shown that sarin simulants can adsorb on metal oxides by forming a bond with metal atoms via a phosphoryl oxygen.
Initially, we observed the DIMP molecule interacting with an Al atom via the O atom of the P--OC$_3$H$_7$ moiety.
To further explore these dynamical effects with our BOMD simulations, Figure \ref{dist1} plots the distance between one of the surface Al atoms and the oxygen of the P--OC$_3$H$_7$ moiety in DIMP at various temperatures. These calculations show that DIMP adsorption occurs within two picoseconds for all temperatures, and the oxygen of the P--OC$_3$H$_7$ moiety of DIMP interacts with the tetra-coordinated Al center. 
%Due to the different initial conditions inherent to BOMD simulations, we observed two different mechanisms where the O3 and O2 oxygen atoms in DIMP interact with the Al48 and Al43 aluminum atoms, respectively, on the alumina surface. 
We did not observe any desorption of DIMP on the alumina surface, as indicated by the small Al48--O3 bond distances in Figure \ref{dist1} (as a side note, the 600$^{\circ}$ C plot in the center of Figure \ref{dist1} does show  dissociation of the Al48--O3 bond, but a new Al11--O2 bond quickly forms thereafter, and the DIMP molecule still remains on the alumina surface). Figure \ref{ads} shows snapshots from our simulations depicting various DIMP adsorption configurations on the alumina surface.
As suggested by prior theoretical calculations,\cite{roman-2019,roman-2020,roman-2021} the adsorption via the O atom of the P--OC$_3$H$_7$ moiety is crucial for propene elimination. As the MD simulations progresses, we also observe interactions between the O (in the --P=O group) and Al atoms.
In short, the DIMP molecule remained adsorbed on the alumina surface in our simulations for all temperatures. 

\begin{figure}
    \centering
    \includegraphics[width=15cm]{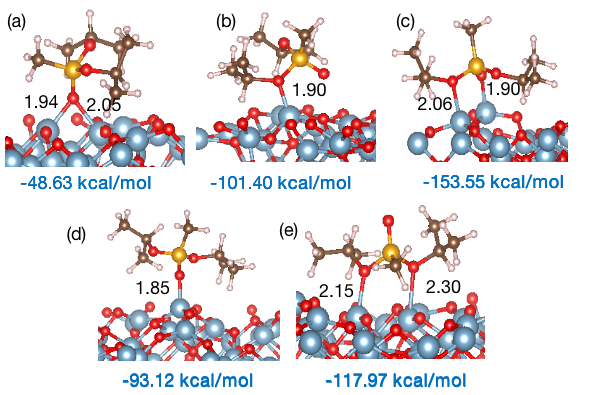}
    \caption{Various adsorption configurations of DIMP on the $\gamma$-Al$_2$O$_3$ (100) surface. Bond distances are shown in angstroms.}
    \label{opt}
\end{figure}

To complement our MetaD simulations, we also calculated adsorption energies, $E_\text{ads}$, using the following expression: 
\begin{equation}
    E_\text{ads} = E_{\text{DIMP} + \text{surface}} - E_{\text{surface}} - E_{\text{DIMP}},
\end{equation}
where $E_\text{surface}$ is the energy of the alumina surface, $E_\text{DIMP}$ is the energy of an isolated gas-phase DIMP molecule, and $E_\text{DIMP +  surface}$ represents the energy of the adsorbed molecule on the surface. A negative value of $E_\text{ads}$ corresponds to an exothermic process and a stable adsorption configuration. The tri-coordinated Al surface atoms are known to be strong Lewis acid-type catalytic sites \cite{wischert-2011} and give large adsorption energies when the DIMP molecule binds to them. Figure \ref{opt} illustrates selected optimized structures of DIMP on the alumina surface. In panel \ref{opt}(a), the DIMP molecule is bonded to two tetra-coordinated Al atoms via the phosphoryl oxygen, resulting in a bridging adsorption with $E_\text{ads} = -48.63$ kcal/mol. In panel \ref{opt}(b), the alkoxy oxygen is bonded to a tri-coordinated Al atom, giving an adsorption energy of -101.40 kcal/mol. The largest adsorption energy (-153.55 kcal/mol) was obtained for the configuration shown in panel \ref{opt}(c) where the alkoxy and phosphoryl oxygen atoms are bonded to two different tri-coordinated Al centers. 
In panel \ref{opt}(d), the phosphoryl oxygen is bonded to a tri-coordinated Al center, giving an adsorption energy of -93.12 kcal/mol. The adsorption energy for configuration in panel \ref{opt}(e) is -117.97 kcal/mol in which the DIMP molecule binds with two tetra-coordinated Al atoms via two alkoxy oxygen atoms. 
Our calculated adsorption energy for the configuration shown in panel 5(a) is in good agreement with a previously studied similar configuration of sarin adsorption on $\gamma$-Al$_2$O$_3$ (-49.2 kcal/mol).\cite{bermudez-2007} 

\subsection{Decomposition of DIMP on the Alumina Surface}
\begin{figure}
    \centering
    \includegraphics[width=12cm]{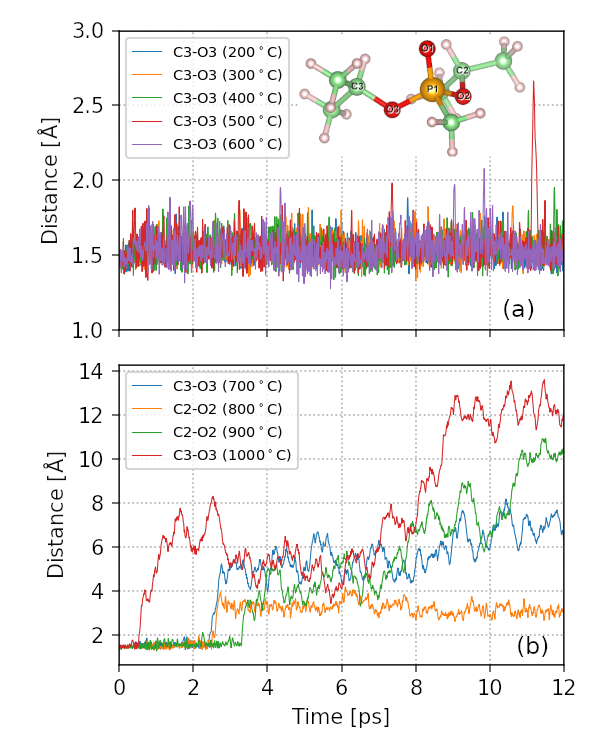}
    \caption{Evolution of C--O distances as a function of time on the alumina surface at various temperatures, which describes DIMP decomposition. The corresponding atom labeling is shown in the inset panel of (a).  }
    \label{dist2}
\end{figure}
We next examine DIMP decomposition mechanisms on the alumina surface. Previous studies have shown that one of the decomposition pathways of DIMP proceeds through the breaking of a C--O bond.\cite{bailin-1975}
%(This is redundant)Therefore, we calculate the fluctuations of the various C-O bonds, and the results are shown in Figure \ref{dist2}.
Figure \ref{dist2}a displays time-dependent variations of C--O bond distances from 200 -- 600$^{\circ}$ C, which fluctuate around 1.50 {\AA} (the covalent C--O bond distance), indicating that the C--O bond remains intact. On the other hand, Figure \ref{dist2}b shows that when the temperature is raised to 700 -- 1000$^{\circ}$ C, the C--O bond lengthens significantly, signifying decomposition. It is worth noting that the timescales for DIMP decomposition varies with temperature, and we observe a very rapid decomposition at 1000$^{\circ}$ C. 

\begin{figure}
    \centering
    \includegraphics[width=15cm]{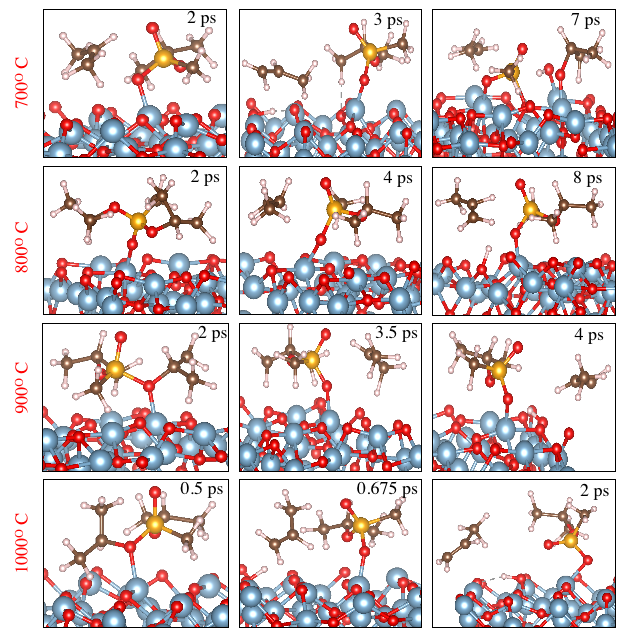}
    \caption{Snapshots of DIMP decomposition from representative MD trajectories at various temperatures.}
    \label{snap}
\end{figure}

Figure \ref{snap} depicts snapshots at crucial points for the decomposition of DIMP  on the alumina surface for a few representative MD trajectories. As mentioned in the previous section, adsorption initially occurs via the oxygen atom of the P--OC$_3$H$_7$ moiety in DIMP and an Al atom on the alumina surface (which occurs at all temperatures).
At 700$^{\circ}$ C, the 3 panels at the top of Figure \ref{snap} depict a possible decomposition pathway in which the C3--O3 bond breaks after 2 ps and an Al--O bond forms with the alumina surface. During this process, a surface-bound n-alkyl (--C$_3$H$_7$) species is formed. At $\sim$3 ps, a single terminal C--H bond from the n-alkyl molecule is broken when a proton is abstracted by an oxygen atom on the alumina surface. This forms propene, an unsaturated compound, as a by-product. As the simulation further proceeds, the P--O bond in DIMP dissociates at 7 ps and forms a new Al--OCHCH$_3$CH$_3$ adsorbate on the alumina surface.  In addition, once the P--O bond in DIMP is cleaved, a new P--O bond is formed between the phosphorus atom and a di-coordinated surface oxygen atom (see the 3 panels at the top of Figure \ref{snap}). At 800 -- 1000$^{\circ}$ C (depicted in the bottom 9 panels of Figure \ref{snap}), DIMP decomposition involves only the propene elimination via a two-step process. The first step of the propene elimination is associated with the dissociation of the C2--O2 (800 -- 900$^{\circ}$ C) or C3--O3 (1000$^{\circ}$ C) bond. The second step comprises the migration of the hydrogen atom from the methyl group of the --C$_3$H$_7$ fragment to one of the surface oxygen atoms. Within these simulated timescales, we did not observe any further decomposition of DIMP within the 800 -- 1000$^{\circ}$ C temperature range.

\subsection{Free energy profiles}
The propene end-products predicted by our BOMD simulations corroborates previous experiments on DIMP, which include thermal decomposition studies, \cite{zegers,senyurt-2020} pyrolysis and combustion in nitrogen/oxygen-rich environments, \cite{yuan-2019} as well as microwave \cite{bailin-1975} and laser-induced \cite{thompson-2019} decomposition under inert environments.
%\textcolor{red}{(a lot of the text about other studies is extraneous and unnecessary Please condense this to only a few sentences and please stop writing unnecessary things like this.)}Zegers and Fisher suggested a two-step mechanism for the thermal decomposition of DIMP.\cite{zegers} According to their proposed mechanism, the first stage commences to the formation of propene and isopropyl methylphosphonate. The second stage involves two competing processes, resulting in propene and methyl phosphonic acid in the first step and methyl(oxo) phosphonuimolate and 2-propanol in the second step. Yuan \textit{et al.} also performed T-jump pyrolysis and combustion of DIMP in nitrogen, air, oxygen-rich environment with a heating rate of $\sim$10$^{\circ}$C/s, $\sim$1800$^{\circ}$C/s, and $\sim$18000$^{\circ}$C/s to temperatures between 500$^{\circ}$ C to 1200$^{\circ}$ C.\cite{yuan-2019} They observed propene, isopropyl methylphosphonate, and methyl phosphonic acid as final decomposition products. The microwave\cite{bailin-1975} and laser-induced\cite{thompson-2019} DIMP decomposition in inert conditions leads to propene as a predominant product. Senyurt \textit{et al.} recently performed thermal decomposition of DIMP, which also demonstrates propene, methyl phosphonic acid, methyl(oxo) phosphonuimolate, and isopropanol as significant products.\cite{senyurt-2020} \textcolor{red}{(it is not necessary to go through the previous studies in this much detail. This really shows that this work lacks focus and substance.)}
Collectively, all of these prior experimental studies detected propene as one of the main by-products of DIMP decomposition, which further supports our BOMD predictions.
\begin{figure}
    \centering
    \includegraphics[width=16cm]{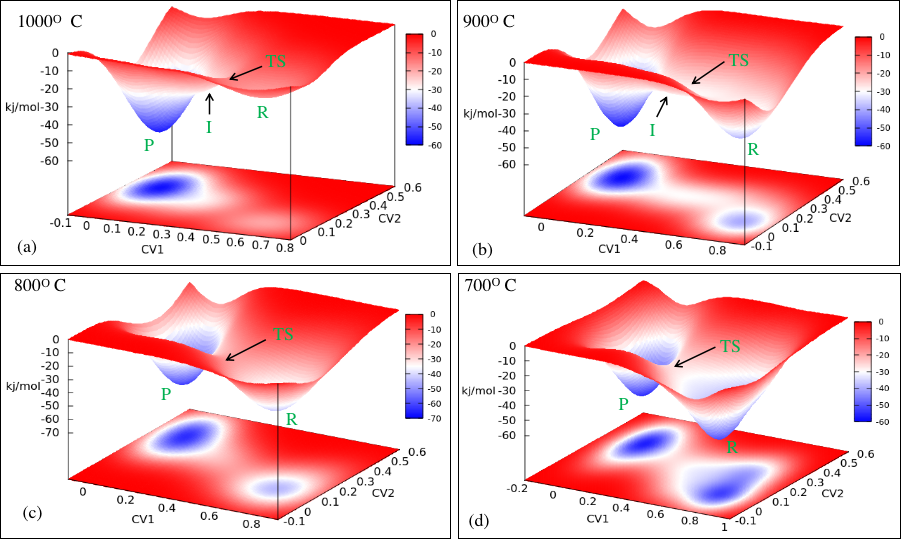}
    \caption{Reconstructed free energy surface for the decomposition of DIMP on the alumina surface at (a) 1000$^{\circ}$ C, (b) 900$^{\circ}$ C, (c) 800$^{\circ}$ C, and (d) 700$^{\circ}$ C. The R, TS, I, and P labels in each free energy surface correspond to the reactant, transition state, intermediate, and product, respectively.}
    \label{free}
\end{figure}
To further investigate the decomposition of DIMP on alumina, we utilized well-tempered metadynamics simulations by adopting two collective variables. We performed two sets of metadynamics simulations (for each temperature within the 200 -- 1000$^{\circ}$ C range) using  different initial conditions. Obtaining accurate free energy profiles from  metadynamics simulations requires (1) longer simulation times until all accessible regions of the potential are explored (with trajectories spanning forward and backward many times between reactant and product states) and (2) a careful selection of collective variables. Our well-tempered metadynamics simulations at various temperatures suggests that when the C--O bond is broken, subsequent reforming of this bond is prohibited (Figure S19), which indicates that the reaction is irreversible. The Supporting Information (Figures S20 and S21) provides further details on the convergence of our free energy profiles between the transition state and reactant basin as a function of time, which demonstrates that our metadynamics simulations are fully converged. 
\begin{figure}
    \centering
    \includegraphics[width=12cm]{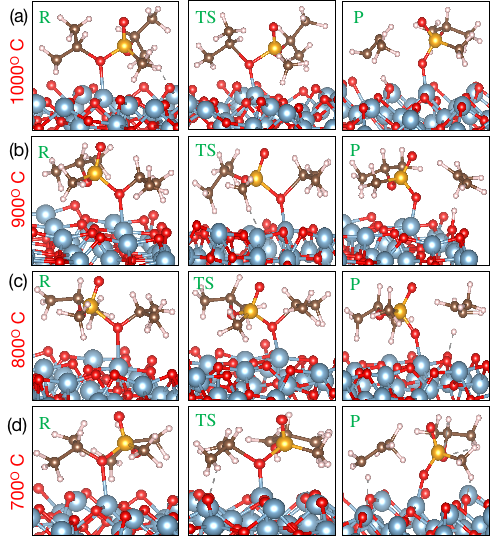}
    \caption{Snapshots from the well-tempered metadynamics simulations within the 700 -- 1000$^{\circ}$ C temperature range. The R, TS, and P labels illustrate the reactant, transition state, and product, respectively.}
    \label{pdt}
\end{figure}

The free energy surface at 1000$^{\circ}$ C is shown in Figure \ref{free} (a).  The two CVs that capture the energetics of DIMP decomposition are presented in Figure S1. In this mechanism, the reactant (R) goes to a product (P) via a transition state (TS) and a small tiny intermediate (I) in between the transition state and product. The reactant state is defined by CV$1=0.75$ (C--O $=1.66$ {\AA}) and CV$2=0.11$ (Al--O $=2.17$ {\AA} ), where DIMP is bonded to the surface Al atom. The decomposed DIMP product, P, is characterized by CV$1=0.03$ (C--O $=3.12$ {\AA}) and CV$2=0.40$ (Al--O $=1.77$ {\AA}), in which the C--O bond of DIMP is cleaved. At this stage, the H atom from the --C$_3$H$_7$ moiety is transferred to a surface oxygen atom, and propene is formed. The C--O bond length stretches to 1.87 {\AA} at the transition state (which has an activation barrier of only 4.11 kJ/mol) and subsequently dissociates.  The reactant, transition state, and product from our well-tempered metadynamics simulations at 1000$^{\circ}$ C are shown in Figure \ref{pdt}(a). In the Supplementary Material, we provide an additional free energy profile (Figure S2) at 1000$^{\circ}$ C that utilizes different initial conditions. For this separate case, the reactant proceeds to the product state via a stable transition state with a tiny minimum, and the activation barrier is $\sim$6.36 kJ/mol. Collectively, the average free energy value from these independent trajectories is 5.24 kJ/mol. 

The reaction mechanism at 900$^{\circ}$ C is very similar to that of 1000$^{\circ}$ C, which also shows a tiny intermediate between the product and transition state, as shown in Figure \ref{free}(b). The free energy barrier value is $\sim$11.85 kJ/mol, and Figure S3 in the Supplementary Material shows the corresponding evolution of CV1 and CV2. The geometries of the reactant, transition state, and product along the metadynamics trajectory are reported in Figure \ref{pdt}(b). We have also performed additional metadynamics calculations using different initial conditions to test the reproducibility of these results. In these additional calculations (depicted in Figure S4 in the Supplementary Information), the reactant proceeds to the product through a single transition state with a free energy barrier value of $\sim$15.73 kJ/mol. The average free energy barrier value obtained from these different initial conditions is 13.79 kJ/mol.

The reconstructed free energy surfaces at 800 and 700$^{\circ}$ C  are shown in Figures \ref{free}(c) and \ref{free}(d), and fluctuations of the corresponding CV values are presented in Figures S5 and S6, respectively. In contrast to the higher temperatures discussed previously, the decomposition of  DIMP at 800 and 700$^{\circ}$ C proceeds via a single transition state. At 800$^{\circ}$ C, the collective variables CV1 (0.71) and CV2 (0.12) define the reactant state R, at which DIMP forms a bond with the surface Al atom (typical C--O and Al--O bond distances at this geometry are 1.53 and 2.25 {\AA}, respectively). The C--O bond in DIMP stretches to 1.91 {\AA} to form a transition state, and the reaction progresses to the product where the C--O bond subsequently dissociates, and propene is formed. 
The net free energy barrier for this activation process is $\sim$22.53 kJ/mol. The additional free energy profile at 800$^{\circ}$ C from the different initial conditions is shown in Figure S7. 
A very similar reaction mechanism is also observed at 700$^{\circ}$ C with a slightly higher activation barrier of $\sim$26.65 kJ/mol. 
Figure S8 in the Supplementary Material represents the topology of another free energy surface at 700$^{\circ}$ C from different initial conditions.
The reactant, transition state, and product geometries from these metadynamics simulations are shown in Figures \ref{pdt}(c) and \ref{pdt}(d) for 800 and 700$^{\circ}$ C, respectively.
%As discussed previously, we carried out additional free energy calculations using different initial conditions shown in the Supplementary Material. Figure S7 plots the other free energy profile at 800$^{\circ}$ C, which shows the reaction proceeds to the product state from the reactant state via transition state with a free energy barrier value of 17.91 KJ/mol. 
%Therefore, the free energy barrier value (obtained from the different initial conditions)  at 800$^{\circ}$ C  is 20.22 KJ/mol. Figure S8 in the Supplementary Material represents the topology of another free energy surface at 700$^{\circ}$ C from the varying initial conditions. Figure S8 exhibits that the DIMP decomposition occurs at two sub-steps. In this mechanism, the reactant (R) first moves to stable intermediate (I) through a transition state 1 (TS1) and then goes to the product (P) state via a transition state 2 (TS2). The maximum reconstructed free energy barrier is determined to be 25.43 KJ/mol. Based on these free energy profiles, the average activation barrier is 26.04 KJ/mol at 700$^{\circ}$ C.
\begin{figure}
    \centering
    \includegraphics[width=12cm]{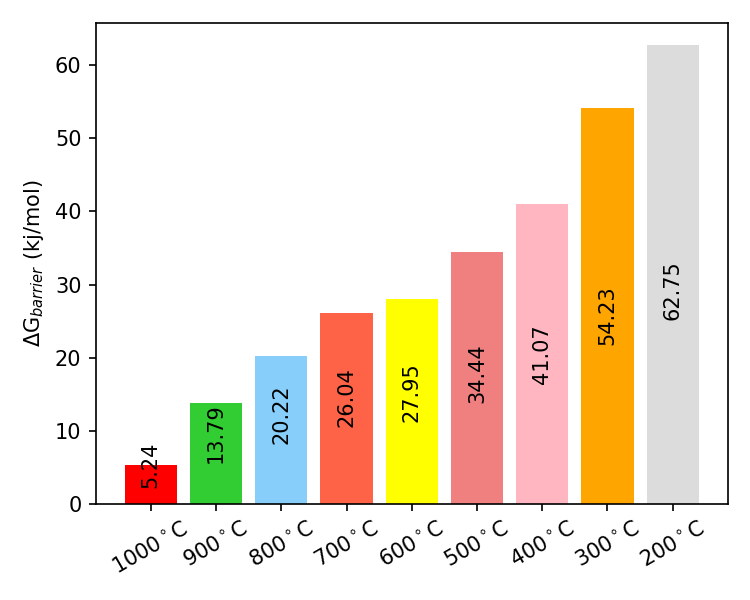}
    \caption{Free energy activation barrier as a function of temperature for the decomposition of DIMP on the alumina surface. The free energy activation barrier value is calculated by averaging over two metadynamics simulations at each temperature.}
    \label{deltag}
\end{figure}
A detailed analysis of free energy profiles for DIMP decomposition on alumina at various temperatures (200 -- 600$^{\circ}$ C) with different initial conditions is given in the Supplementary Material (Figures S9-S18). Figure \ref{deltag} summarizes the free energy activation barrier as a function of temperature, which shows that the free energy activation barrier decreases with temperature.
Static DFT calculations have obtained free energy activation barrier values of 113 kJ/mol (ZnO surface),\cite{roman-2019} 108.0 kJ/mol (rutile surface),\cite{roman-2021} 122.6 kJ/mol (anatase surface),\cite{roman-2021} and 53.7 kJ/mol (MoO$_2$ surface).\cite{roman-2020} However, the free energy activation barriers for C--O bond breaking from our  simulations are much lower than these previously reported values.
We also note that the free energy activation barrier value at 600$^{\circ}$ C is very similar to that of 700$^{\circ}$ C. 

\begin{figure}
    \centering
    \includegraphics[width=16cm]{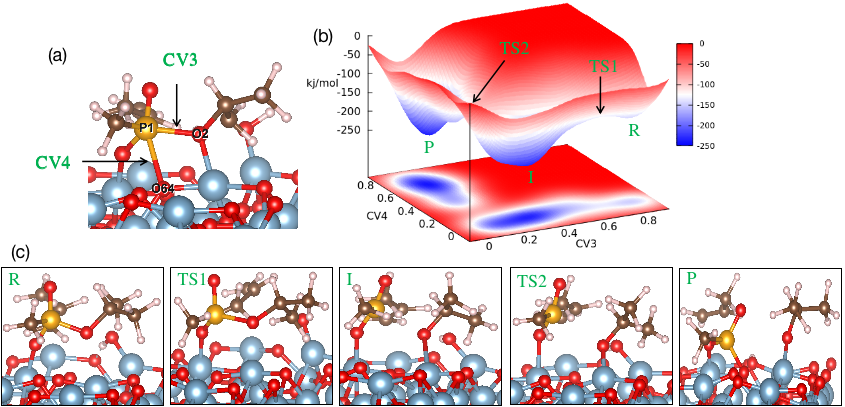}
    \caption{Panel (a) depicts collective variables used in studying the formation of an Al--OCHCH$_3$CH$_3$ adsorbate at 700$^{\circ}$ C, and panel (b) shows the 3D reconstructed free energy surface, where R, TS1, I, TS2, and P correspond to the reactant, transition state 1, intermediate, transition state 2, and product, respectively. Panel (c) shows snapshots of R, TS1, I, TS2, and P along the metadynamics trajectory.}
    \label{free2}
\end{figure}
At 700$^{\circ}$ C, we observed formation of an Al--OCHCH$_3$CH$_3$ adsorbate in addition to propene. To further analyze these energetics with metadynamics simulations, we utilized two collective variables: CV3 [P1--O2] and CV4 [P1--064], which denote the coordination number of the P1 phosphorus with respect to the O2 and O64 oxygen atoms, respectively, in the DIMP molecule. The CVs are shown in Figure \ref{free2}(a), and the corresponding free energy surface is depicted in Figure \ref{free2}(b). The reaction mechanism takes place in two substeps, as shown in Figure \ref{free2}(c). The reactant is described by CV$3= 0.80$ and CV$4= 0.05$, in which the P1--O2 and P1--O64 bond distances are 1.69 and 3.02 {\AA}, respectively. The P1--O2 bond first stretches to 1.95 {\AA} and forms the first transition state (TS1). The P1--O2 bond then dissociates to 2.46 and creates a stable intermediate, I (CV$3= 0.15$, CV$4 = 0.04$), giving a free energy activation of 18.51 kJ/mol. The reaction proceeds over a second transition state (TS2) in which the P1--O2 bond is further stretched and the resulting --OCHCH$_3$CH$_3$ fragment binds to the surface Al atom. In the product state, the distance from the P1 atom to the surface oxygen (O64) shortens (1.74 {\AA}), indicating the formation of a P1--O64 covalent bond. In contrast, the P1--O2 bond length further increases, showing the complete decomposition of the DIMP molecule.

Collectively, our AIMD calculations show that the decomposition of DIMP most likely progresses via a propene elimination on the alumina surface.
The final step of the propene elimination occurs via the abstraction of a  hydrogen atom by the surface oxygen atom of gamma-alumina. We also performed metadynamics simulations to evaluate the free energy activation of this process. The C-H coordination (CV5) and H--O coordination (CV6) were selected as collective variables, as shown in Figure S22 in the Supporting Information. A Gaussian with a height of 0.0001 Hartree was used for these simulations. Figure S23 in the Supporting Information illustrates free energy profiles for the C--H activation within the 200 -- 500$^{\circ}$ C temperature range. In Figure S24, we show free energy profiles in the 600 -- 900$^{\circ}$ C temperature range. The free energy profile at 1000$^{\circ}$ C is shown in Figure S25a, and Figure S25b summarizes the free energy activation barrier as a function of temperature. The free energy activation barriers range from 3.91 -- 1.29 kJ/mol for the 200 -- 1000$^{\circ}$ C temperature range.
We find that the free energy activation barrier value decreases with increasing the temperature. The small free energy barrier value in the 600 -- 1000$^{\circ}$ C temperature range lies between 5 -- 28 kJ/mol, which can be easily accessible by experiment. Our theoretical findings are also consistent with experimental studies, which also identified propene as the main gas-phase product of the DIMP decomposition. In summary, our AIMD simulations show that the $\gamma$-Al$_2$O$_3$ surface can trap and subsequently decompose DIMP due to strong electrostatic attractions between the phosphoryl oxygen and surface Al atoms.

\section{Conclusions}
In this work, we have harnessed large-scale \textit{ab initio} molecular dynamics calculations to investigate the adsorption and decomposition of DIMP on the $\gamma$-alumina surface over a wide range of temperatures.
%Redundant: In this research, we explored the adsorption and decomposition of DIMP on the alumina surface.
%In particular, we found that the alumina surface effectively adsorbs and decomposes the DIMP molecule.
Our DFT-based molecular dynamics calculations predict a spontaneous decomposition of DIMP with propene as the main by-product within the 700 -- 1000$^{\circ}$ C temperature range (the decomposition reaction leads to propene and an Al--OCHCH$_3$CH$_3$ adsorbate at 700$^{\circ}$ C). 
Due to the short-time scales inherent to BOMD simulations, it is likely that a similar decomposition would also occur at lower temperatures but would take longer to happen.
Well-tempered metadynamics AIMD simulations were performed for temperatures ranging from 200 -- 1000$^{\circ}$ C to provide atomistic-level details of the reaction path and associated energetics of DIMP decomposition. 
Our metadynamics calculations also reveal that the free energy barrier value decreases with temperature. 
The low free energy barrier values at higher temperatures suggest that that reaction is extremely fast at higher temperatures and is likely to occur at a lower rate at lower temperatures.

In our study, we obtained free energy values of 62.75 to 5.24 kJ/mol within the 200 -- 1000$^{\circ}$ C temperature range. Recent experiments on CWAs have also been carried out under similar temperatures, including vapor phase decomposition of DMMP (a sarin surrogate) at 500 -- 800 K,\cite{MUKHOPADHYAY2021} sarin decomposition on  TiO$_2$ nanoparticles at 1000 K,\cite{roman-2021} and decomposition of CWAs at 2000 K.\cite{senyurt-2021,neupane-2018, mathieu-2019}
%The calculated free energy values are much lower than those obtained from the static DFT calculations on various metal-oxide surfaces.\cite{roman-2019,roman-2020,roman-2021} 
%Recently, the vapor phase decomposition of DMMP (a sarin surrogate) on a metal oxide surface was observed in a flow reactor experiment where temperatures of 500-800 K were achieved within a residence time range of sub-seconds to several seconds.\cite{MUKHOPADHYAY2021} Another study showed sarin decomposition on  TiO$_2$ nanoparticles where the sample was heated up to 1000 K.\cite{roman-2021} Recently, the rapid decomposition of CWAs was studied using shock heating at temperatures as high as 2000 K.\cite{senyurt-2021,neupane-2018, mathieu-2019} 
Collectively, all of these experiments showed that the temperature ranges studied in this work can also be achieved under operational conditions, and DIMP would decompose on the $\gamma$-Al$_2$O$_3$ surface. Due to the structural similarity between DIMP and sarin, our calculations provide additional insight into decomposition mechanisms of both these molecules and elucidate atomic details of sarin decomposition on candidate metal-oxide surfaces.
As a final remark, this study serves as a convincing demonstration of the use of DFT-based molecular dynamics simulations for investigating the interactions of CWAs with existing metal-oxides, which can be used to guide experimental efforts on these hazardous compounds.

\begin{acknowledgement}
The project or effort depicted was or is sponsored by the Department of the Defense, Defense Threat Reduction Agency under the Materials Science in Extreme Environments University Research Alliance, HDTRA1-20-2-0001. The content of the information does not necessarily reflect the position or the policy of the federal government, and no official endorsement should be inferred. 
\end{acknowledgement}

\begin{suppinfo}
Additional materials on free energy profiles, fluctuations of collective variables, convergence tests of metadynamics calculations, and fluctuations of the C--O bond distance in DIMP from metadynamics calculations. 
\end{suppinfo}

\bibliography{acs-achemso}

\providecommand{\latin}[1]{#1}
\makeatletter
\providecommand{\doi}
  {\begingroup\let\do\@makeother\dospecials
  \catcode`\{=1 \catcode`\}=2 \doi@aux}
\providecommand{\doi@aux}[1]{\endgroup\texttt{#1}}
\makeatother
\providecommand*\mcitethebibliography{\thebibliography}
\csname @ifundefined\endcsname{endmcitethebibliography}
  {\let\endmcitethebibliography\endthebibliography}{}
\begin{mcitethebibliography}{61}
\providecommand*\natexlab[1]{#1}
\providecommand*\mciteSetBstSublistMode[1]{}
\providecommand*\mciteSetBstMaxWidthForm[2]{}
\providecommand*\mciteBstWouldAddEndPuncttrue
  {\def\EndOfBibitem{\unskip.}}
\providecommand*\mciteBstWouldAddEndPunctfalse
  {\let\EndOfBibitem\relax}
\providecommand*\mciteSetBstMidEndSepPunct[3]{}
\providecommand*\mciteSetBstSublistLabelBeginEnd[3]{}
\providecommand*\EndOfBibitem{}
\mciteSetBstSublistMode{f}
\mciteSetBstMaxWidthForm{subitem}{(\alph{mcitesubitemcount})}
\mciteSetBstSublistLabelBeginEnd
  {\mcitemaxwidthsubitemform\space}
  {\relax}
  {\relax}

\bibitem[Kim \latin{et~al.}(2011)Kim, Tsay, Atwood, and Churchill]{kim-2011}
Kim,~K.; Tsay,~O.~G.; Atwood,~D.~A.; Churchill,~D.~G. Destruction and Detection
  of Chemical Warfare Agents. \emph{Chem. Rev.} \textbf{2011}, \emph{111},
  5345--5403\relax
\mciteBstWouldAddEndPuncttrue
\mciteSetBstMidEndSepPunct{\mcitedefaultmidpunct}
{\mcitedefaultendpunct}{\mcitedefaultseppunct}\relax
\EndOfBibitem
\bibitem[Picard \latin{et~al.}(2019)Picard, Chataigner, Maddaluno, and
  Legros]{picard-2019}
Picard,~B.; Chataigner,~I.; Maddaluno,~J.; Legros,~J. Introduction to chemical
  warfare agents{,} relevant simulants and modern neutralisation methods.
  \emph{Org. Biomol. Chem.} \textbf{2019}, \emph{17}, 6528--6537\relax
\mciteBstWouldAddEndPuncttrue
\mciteSetBstMidEndSepPunct{\mcitedefaultmidpunct}
{\mcitedefaultendpunct}{\mcitedefaultseppunct}\relax
\EndOfBibitem
\bibitem[Michalkova \latin{et~al.}(2007)Michalkova, Paukku, Majumdar, and
  Leszczynski]{michalkova-2007}
Michalkova,~A.; Paukku,~Y.; Majumdar,~D.; Leszczynski,~J. Theoretical study of
  adsorption of tabun on calcium oxide clusters. \emph{Chem. Phys. Lett.}
  \textbf{2007}, \emph{438}, 72--77\relax
\mciteBstWouldAddEndPuncttrue
\mciteSetBstMidEndSepPunct{\mcitedefaultmidpunct}
{\mcitedefaultendpunct}{\mcitedefaultseppunct}\relax
\EndOfBibitem
\bibitem[Wagner \latin{et~al.}(2000)Wagner, Koper, Lucas, Decker, and
  Klabunde]{wagner-2000}
Wagner,~G.~W.; Koper,~O.~B.; Lucas,~E.; Decker,~S.; Klabunde,~K.~J. Reactions
  of VX, GD, and HD with Nanosize CaO: Autocatalytic Dehydrohalogenation of HD.
  \emph{J. Phys. Chem. B} \textbf{2000}, \emph{104}, 5118--5123\relax
\mciteBstWouldAddEndPuncttrue
\mciteSetBstMidEndSepPunct{\mcitedefaultmidpunct}
{\mcitedefaultendpunct}{\mcitedefaultseppunct}\relax
\EndOfBibitem
\bibitem[Michalkova \latin{et~al.}(2004)Michalkova, Ilchenko, Gorb, and
  Leszczynski]{michalkova-2004}
Michalkova,~A.; Ilchenko,~M.; Gorb,~L.; Leszczynski,~J. Theoretical Study of
  the Adsorption and Decomposition of Sarin on Magnesium Oxide. \emph{J. Phys.
  Chem. B} \textbf{2004}, \emph{108}, 5294--5303\relax
\mciteBstWouldAddEndPuncttrue
\mciteSetBstMidEndSepPunct{\mcitedefaultmidpunct}
{\mcitedefaultendpunct}{\mcitedefaultseppunct}\relax
\EndOfBibitem
\bibitem[Wagner \latin{et~al.}(1999)Wagner, Bartram, Koper, and
  Klabunde]{wagner-1999}
Wagner,~G.~W.; Bartram,~P.~W.; Koper,~O.; Klabunde,~K.~J. Reactions of VX, GD,
  and HD with Nanosize MgO. \emph{J. Phys. Chem. B} \textbf{1999}, \emph{103},
  3225--3228\relax
\mciteBstWouldAddEndPuncttrue
\mciteSetBstMidEndSepPunct{\mcitedefaultmidpunct}
{\mcitedefaultendpunct}{\mcitedefaultseppunct}\relax
\EndOfBibitem
\bibitem[Mahato \latin{et~al.}(2009)Mahato, Prasad, Singh, Acharya, Srivastava,
  and Vijayaraghavan]{mahato-2009}
Mahato,~T.; Prasad,~G.; Singh,~B.; Acharya,~J.; Srivastava,~A.;
  Vijayaraghavan,~R. Nanocrystalline zinc oxide for the decontamination of
  sarin. \emph{J. Hazard. Mater.} \textbf{2009}, \emph{165}, 928--932\relax
\mciteBstWouldAddEndPuncttrue
\mciteSetBstMidEndSepPunct{\mcitedefaultmidpunct}
{\mcitedefaultendpunct}{\mcitedefaultseppunct}\relax
\EndOfBibitem
\bibitem[Prasad \latin{et~al.}(2007)Prasad, Mahato, Singh, Ganesan, Pandey, and
  Sekhar]{prasad-2007}
Prasad,~G.; Mahato,~T.; Singh,~B.; Ganesan,~K.; Pandey,~P.; Sekhar,~K.
  Detoxification reactions of sulphur mustard on the surface of zinc oxide
  nanosized rods. \emph{J. Hazard. Mater.} \textbf{2007}, \emph{149},
  460--464\relax
\mciteBstWouldAddEndPuncttrue
\mciteSetBstMidEndSepPunct{\mcitedefaultmidpunct}
{\mcitedefaultendpunct}{\mcitedefaultseppunct}\relax
\EndOfBibitem
\bibitem[Prasad \latin{et~al.}(2011)Prasad, Ramacharyulu, Singh, Batra,
  Srivastava, Ganesan, and Vijayaraghavan]{prasad-2011}
Prasad,~G.; Ramacharyulu,~P.; Singh,~B.; Batra,~K.; Srivastava,~A.~R.;
  Ganesan,~K.; Vijayaraghavan,~R. Sun light assisted photocatalytic
  decontamination of sulfur mustard using ZnO nanoparticles. \emph{J. Mol.
  Catal. A: Chem.} \textbf{2011}, \emph{349}, 55--62\relax
\mciteBstWouldAddEndPuncttrue
\mciteSetBstMidEndSepPunct{\mcitedefaultmidpunct}
{\mcitedefaultendpunct}{\mcitedefaultseppunct}\relax
\EndOfBibitem
\bibitem[Hirakawa \latin{et~al.}(2010)Hirakawa, Sato, Komano, Kishi, Nishimoto,
  Mera, Kugishima, Sano, Ichinose, Negishi, Seto, and Takeuchi]{hirakawa-2010}
Hirakawa,~T.; Sato,~K.; Komano,~A.; Kishi,~S.; Nishimoto,~C.~K.; Mera,~N.;
  Kugishima,~M.; Sano,~T.; Ichinose,~H.; Negishi,~N.; Seto,~Y.; Takeuchi,~K.
  Experimental Study on Adsorption and Photocatalytic Decomposition of
  Isopropyl Methylphosphonofluoridate at Surface of TiO$_2$ Photocatalyst.
  \emph{J. Phys. Chem. C} \textbf{2010}, \emph{114}, 2305--2314\relax
\mciteBstWouldAddEndPuncttrue
\mciteSetBstMidEndSepPunct{\mcitedefaultmidpunct}
{\mcitedefaultendpunct}{\mcitedefaultseppunct}\relax
\EndOfBibitem
\bibitem[Prasad \latin{et~al.}(2008)Prasad, Mahato, Singh, Ganesan, Srivastava,
  Kaushik, and Vijayraghavan]{prasad-2008}
Prasad,~G.~K.; Mahato,~T.~H.; Singh,~B.; Ganesan,~K.; Srivastava,~A.~R.;
  Kaushik,~M.~P.; Vijayraghavan,~R. Decontamination of sulfur mustard and sarin
  on titania nanotubes. \emph{AIChE Journal} \textbf{2008}, \emph{54},
  2957--2963\relax
\mciteBstWouldAddEndPuncttrue
\mciteSetBstMidEndSepPunct{\mcitedefaultmidpunct}
{\mcitedefaultendpunct}{\mcitedefaultseppunct}\relax
\EndOfBibitem
\bibitem[Prasad \latin{et~al.}(2009)Prasad, Singh, Ganesan, Batra, Kumeria,
  Gutch, and Vijayaraghavan]{prasad-2009}
Prasad,~G.; Singh,~B.; Ganesan,~K.; Batra,~A.; Kumeria,~T.; Gutch,~P.;
  Vijayaraghavan,~R. Modified titania nanotubes for decontamination of sulphur
  mustard. \emph{J. Hazard. Mater.} \textbf{2009}, \emph{167}, 1192--1197\relax
\mciteBstWouldAddEndPuncttrue
\mciteSetBstMidEndSepPunct{\mcitedefaultmidpunct}
{\mcitedefaultendpunct}{\mcitedefaultseppunct}\relax
\EndOfBibitem
\bibitem[Ramacharyulu \latin{et~al.}(2012)Ramacharyulu, Prasad, Ganesan, and
  Singh]{rama-2012}
Ramacharyulu,~P.; Prasad,~G.; Ganesan,~K.; Singh,~B. Photocatalytic
  decontamination of sulfur mustard using titania nanomaterials. \emph{J. Mol.
  Catal. A: Chem.} \textbf{2012}, \emph{353-354}, 132--137\relax
\mciteBstWouldAddEndPuncttrue
\mciteSetBstMidEndSepPunct{\mcitedefaultmidpunct}
{\mcitedefaultendpunct}{\mcitedefaultseppunct}\relax
\EndOfBibitem
\bibitem[Sato \latin{et~al.}(2011)Sato, Hirakawa, Komano, Kishi, Nishimoto,
  Mera, Kugishima, Sano, Ichinose, Negishi, Seto, and Takeuchi]{sato-2011}
Sato,~K.; Hirakawa,~T.; Komano,~A.; Kishi,~S.; Nishimoto,~C.~K.; Mera,~N.;
  Kugishima,~M.; Sano,~T.; Ichinose,~H.; Negishi,~N.; Seto,~Y.; Takeuchi,~K.
  Titanium dioxide photocatalysis to decompose isopropyl
  methylphosphonofluoridate (GB) in gas phase. \emph{Appl. Catal. B: Environ.}
  \textbf{2011}, \emph{106}, 316--322\relax
\mciteBstWouldAddEndPuncttrue
\mciteSetBstMidEndSepPunct{\mcitedefaultmidpunct}
{\mcitedefaultendpunct}{\mcitedefaultseppunct}\relax
\EndOfBibitem
\bibitem[Štengl \latin{et~al.}(2012)Štengl, Grygar, Opluštil, and
  Němec]{STENGL-2012}
Štengl,~V.; Grygar,~T.~M.; Opluštil,~F.; Němec,~T. Ge4+ doped TiO$_2$ for
  stoichiometric degradation of warfare agents. \emph{J. Hazard. Mater.}
  \textbf{2012}, \emph{227-228}, 62--67\relax
\mciteBstWouldAddEndPuncttrue
\mciteSetBstMidEndSepPunct{\mcitedefaultmidpunct}
{\mcitedefaultendpunct}{\mcitedefaultseppunct}\relax
\EndOfBibitem
\bibitem[Wagner \latin{et~al.}(2008)Wagner, Chen, and Wu]{wagner-2008}
Wagner,~G.~W.; Chen,~Q.; Wu,~Y. Reactions of VX, GD, and HD with Nanotubular
  Titania. \emph{J. Phys. Chem. C} \textbf{2008}, \emph{112},
  11901--11906\relax
\mciteBstWouldAddEndPuncttrue
\mciteSetBstMidEndSepPunct{\mcitedefaultmidpunct}
{\mcitedefaultendpunct}{\mcitedefaultseppunct}\relax
\EndOfBibitem
\bibitem[Kuiper \latin{et~al.}(1976)Kuiper, {van Bokhoven}, and
  Medema]{kuiper-1976}
Kuiper,~A.; {van Bokhoven},~J.; Medema,~J. The role of heterogeneity in the
  kinetics of a surface reaction: I. Infrared characterization of the
  adsorption structures of organophosphonates and their decomposition. \emph{J.
  Catal.} \textbf{1976}, \emph{43}, 154--167\relax
\mciteBstWouldAddEndPuncttrue
\mciteSetBstMidEndSepPunct{\mcitedefaultmidpunct}
{\mcitedefaultendpunct}{\mcitedefaultseppunct}\relax
\EndOfBibitem
\bibitem[Saxena \latin{et~al.}(2009)Saxena, Sharma, Srivastava, Singh, Gutch,
  and Semwal]{saxena-2009}
Saxena,~A.; Sharma,~A.; Srivastava,~A.~K.; Singh,~B.; Gutch,~P.~K.;
  Semwal,~R.~P. Kinetics of adsorption of sulfur mustard on Al$_2$O$_3$
  nanoparticles with and without impregnants. \emph{J. Chem. Technol.
  Biotechnol.} \textbf{2009}, \emph{84}, 1860--1872\relax
\mciteBstWouldAddEndPuncttrue
\mciteSetBstMidEndSepPunct{\mcitedefaultmidpunct}
{\mcitedefaultendpunct}{\mcitedefaultseppunct}\relax
\EndOfBibitem
\bibitem[Saxena \latin{et~al.}(2010)Saxena, Srivastava, Singh, Gupta,
  Suryanarayana, and Pandey]{saxena-2010}
Saxena,~A.; Srivastava,~A.~K.; Singh,~B.; Gupta,~A.~K.; Suryanarayana,~M.~V.;
  Pandey,~P. Kinetics of adsorptive removal of DEClP and GB on impregnated
  Al$_2$O$_3$ nanoparticles. \emph{J. Hazard. Mater.} \textbf{2010},
  \emph{175}, 795--801\relax
\mciteBstWouldAddEndPuncttrue
\mciteSetBstMidEndSepPunct{\mcitedefaultmidpunct}
{\mcitedefaultendpunct}{\mcitedefaultseppunct}\relax
\EndOfBibitem
\bibitem[Wagner \latin{et~al.}(2001)Wagner, Procell, O'Connor, Munavalli,
  Carnes, Kapoor, and Klabunde]{wagner-2001}
Wagner,~G.~W.; Procell,~L.~R.; O'Connor,~R.~J.; Munavalli,~S.; Carnes,~C.~L.;
  Kapoor,~P.~N.; Klabunde,~K.~J. Reactions of VX, GB, GD, and HD with Nanosize
  Al$_2$O$_3$. Formation of Aluminophosphonates. \emph{J. Am. Chem. Soc.}
  \textbf{2001}, \emph{123}, 1636--1644\relax
\mciteBstWouldAddEndPuncttrue
\mciteSetBstMidEndSepPunct{\mcitedefaultmidpunct}
{\mcitedefaultendpunct}{\mcitedefaultseppunct}\relax
\EndOfBibitem
\bibitem[Walenta \latin{et~al.}(2020)Walenta, Xu, Tesvara, O’Connor, Sautet,
  and Friend]{walenta-2020}
Walenta,~C.~A.; Xu,~F.; Tesvara,~C.; O’Connor,~C.~R.; Sautet,~P.;
  Friend,~C.~M. Facile Decomposition of Organophosphonates by Dual Lewis Sites
  on a Fe$_3$O$_4$(111) Film. \emph{J. Phys. Chem. C} \textbf{2020},
  \emph{124}, 12432--12441\relax
\mciteBstWouldAddEndPuncttrue
\mciteSetBstMidEndSepPunct{\mcitedefaultmidpunct}
{\mcitedefaultendpunct}{\mcitedefaultseppunct}\relax
\EndOfBibitem
\bibitem[Trotochaud \latin{et~al.}(2017)Trotochaud, Tsyshevsky, Holdren, Fears,
  Head, Yu, Karslıoğlu, Pletincx, Eichhorn, Owrutsky, Long, Zachariah,
  Kuklja, and Bluhm]{troto-2017}
Trotochaud,~L.; Tsyshevsky,~R.; Holdren,~S.; Fears,~K.; Head,~A.~R.; Yu,~Y.;
  Karslıoğlu,~O.; Pletincx,~S.; Eichhorn,~B.; Owrutsky,~J.; Long,~J.;
  Zachariah,~M.; Kuklja,~M.~M.; Bluhm,~H. Spectroscopic and Computational
  Investigation of Room-Temperature Decomposition of a Chemical Warfare Agent
  Simulant on Polycrystalline Cupric Oxide. \emph{Chem. Mater.} \textbf{2017},
  \emph{29}, 7483--7496\relax
\mciteBstWouldAddEndPuncttrue
\mciteSetBstMidEndSepPunct{\mcitedefaultmidpunct}
{\mcitedefaultendpunct}{\mcitedefaultseppunct}\relax
\EndOfBibitem
\bibitem[Digne \latin{et~al.}(2004)Digne, Sautet, Raybaud, Euzen, and
  Toulhoat]{dinge-2004}
Digne,~M.; Sautet,~P.; Raybaud,~P.; Euzen,~P.; Toulhoat,~H. Use of DFT to
  achieve a rational understanding of acid–basic properties of
  $\gamma$-alumina surfaces. \emph{J. Catal.} \textbf{2004}, \emph{226},
  54--68\relax
\mciteBstWouldAddEndPuncttrue
\mciteSetBstMidEndSepPunct{\mcitedefaultmidpunct}
{\mcitedefaultendpunct}{\mcitedefaultseppunct}\relax
\EndOfBibitem
\bibitem[Gu \latin{et~al.}(2018)Gu, Wang, and Leszczynski]{gu-2018}
Gu,~J.; Wang,~J.; Leszczynski,~J. Structure and Energetics of (111) Surface of
  $\gamma$-Al$_2$O$_3$: Insights from DFT Including Periodic Boundary Approach.
  \emph{ACS Omega} \textbf{2018}, \emph{3}, 1881--1888\relax
\mciteBstWouldAddEndPuncttrue
\mciteSetBstMidEndSepPunct{\mcitedefaultmidpunct}
{\mcitedefaultendpunct}{\mcitedefaultseppunct}\relax
\EndOfBibitem
\bibitem[Roy \latin{et~al.}(2012)Roy, Mpourmpakis, Hong, Vlachos, Bhan, and
  Gorte]{roy-2012}
Roy,~S.; Mpourmpakis,~G.; Hong,~D.-Y.; Vlachos,~D.~G.; Bhan,~A.; Gorte,~R.~J.
  Mechanistic Study of Alcohol Dehydration on $\gamma$-Al$_2$O$_3$. \emph{ACS
  Catal.} \textbf{2012}, \emph{2}, 1846--1853\relax
\mciteBstWouldAddEndPuncttrue
\mciteSetBstMidEndSepPunct{\mcitedefaultmidpunct}
{\mcitedefaultendpunct}{\mcitedefaultseppunct}\relax
\EndOfBibitem
\bibitem[Christiansen \latin{et~al.}(2013)Christiansen, Mpourmpakis, and
  Vlachos]{christ-2013}
Christiansen,~M.~A.; Mpourmpakis,~G.; Vlachos,~D.~G. Density Functional
  Theory-Computed Mechanisms of Ethylene and Diethyl Ether Formation from
  Ethanol on $\gamma$-Al$_2$O$_3$(100). \emph{ACS Catal.} \textbf{2013},
  \emph{3}, 1965--1975\relax
\mciteBstWouldAddEndPuncttrue
\mciteSetBstMidEndSepPunct{\mcitedefaultmidpunct}
{\mcitedefaultendpunct}{\mcitedefaultseppunct}\relax
\EndOfBibitem
\bibitem[Bermudez(2007)]{bermudez-2007}
Bermudez,~V.~M. Quantum-Chemical Study of the Adsorption of DMMP and Sarin on
  $\gamma$-Al$_2$O$_3$. \emph{J. Phys. Chem. C} \textbf{2007}, \emph{111},
  3719--3728\relax
\mciteBstWouldAddEndPuncttrue
\mciteSetBstMidEndSepPunct{\mcitedefaultmidpunct}
{\mcitedefaultendpunct}{\mcitedefaultseppunct}\relax
\EndOfBibitem
\bibitem[Bailin \latin{et~al.}(1975)Bailin, Sibert, Jonas, and
  Bell]{bailin-1975}
Bailin,~L.~J.; Sibert,~M.~E.; Jonas,~L.~A.; Bell,~A.~T. Microwave decomposition
  of toxic vapor simulants. \emph{Environ. Sci. Technol.} \textbf{1975},
  \emph{9}, 254--258\relax
\mciteBstWouldAddEndPuncttrue
\mciteSetBstMidEndSepPunct{\mcitedefaultmidpunct}
{\mcitedefaultendpunct}{\mcitedefaultseppunct}\relax
\EndOfBibitem
\bibitem[Gibson and Sibener(2018)Gibson, and Sibener]{gibson-2018}
Gibson,~K.~D.; Sibener,~S.~J. Fate of Some Chemical Warfare Simulants Adsorbed
  on an Inert Surface when Exposed to Rapid Laser Initiated Heating. \emph{J.
  Phys. Chem. C} \textbf{2018}, \emph{122}, 24684--24689\relax
\mciteBstWouldAddEndPuncttrue
\mciteSetBstMidEndSepPunct{\mcitedefaultmidpunct}
{\mcitedefaultendpunct}{\mcitedefaultseppunct}\relax
\EndOfBibitem
\bibitem[Thompson \latin{et~al.}(2019)Thompson, Brann, Purdy, Graham, McMillan,
  and Sibener]{thompson-2019}
Thompson,~R.~S.; Brann,~M.~R.; Purdy,~E.~H.; Graham,~J.~D.; McMillan,~A.~A.;
  Sibener,~S.~J. Rapid Laser-Induced Temperature Jump Decomposition of the
  Nerve Agent Simulant Diisopropyl Methylphosphonate under Atmospheric
  Conditions. \emph{J. Phys. Chem. C} \textbf{2019}, \emph{123},
  21564--21570\relax
\mciteBstWouldAddEndPuncttrue
\mciteSetBstMidEndSepPunct{\mcitedefaultmidpunct}
{\mcitedefaultendpunct}{\mcitedefaultseppunct}\relax
\EndOfBibitem
\bibitem[Senyurt \latin{et~al.}(2021)Senyurt, Schoenitz, and
  Dreizin]{senyurt-2021}
Senyurt,~E.~I.; Schoenitz,~M.; Dreizin,~E.~L. Rapid destruction of sarin
  surrogates by gas phase reactions with focus on diisopropyl methylphosphonate
  (DIMP). \emph{Def. Technol.} \textbf{2021}, \emph{17}, 703--714\relax
\mciteBstWouldAddEndPuncttrue
\mciteSetBstMidEndSepPunct{\mcitedefaultmidpunct}
{\mcitedefaultendpunct}{\mcitedefaultseppunct}\relax
\EndOfBibitem
\bibitem[Glaude \latin{et~al.}(2002)Glaude, Melius, Pitz, and
  Westbrook]{glaude-2002}
Glaude,~P.; Melius,~C.; Pitz,~W.; Westbrook,~C. Detailed chemical kinetic
  reaction mechanisms for incineration of organophosphorus and
  fluoroorganophosphorus compounds. \emph{Proc. Combust. Inst.} \textbf{2002},
  \emph{29}, 2469--2476\relax
\mciteBstWouldAddEndPuncttrue
\mciteSetBstMidEndSepPunct{\mcitedefaultmidpunct}
{\mcitedefaultendpunct}{\mcitedefaultseppunct}\relax
\EndOfBibitem
\bibitem[{VandeVondele} \latin{et~al.}(2005){VandeVondele}, {Krack}, {Mohamed},
  {Parrinello}, {Chassaing}, and {Hutter}]{joost_cpc_2005}
{VandeVondele},~J.; {Krack},~M.; {Mohamed},~F.; {Parrinello},~M.;
  {Chassaing},~T.; {Hutter},~J. {QUICKSTEP: Fast and accurate density
  functional calculations using a mixed Gaussian and plane waves approach}.
  \emph{Comput. Phys. Commun.} \textbf{2005}, \emph{167}, 103--128\relax
\mciteBstWouldAddEndPuncttrue
\mciteSetBstMidEndSepPunct{\mcitedefaultmidpunct}
{\mcitedefaultendpunct}{\mcitedefaultseppunct}\relax
\EndOfBibitem
\bibitem[Perdew \latin{et~al.}(1996)Perdew, Burke, and Ernzerhof]{perdew-1996}
Perdew,~J.~P.; Burke,~K.; Ernzerhof,~M. Generalized Gradient Approximation Made
  Simple. \emph{Phys. Rev. Lett.} \textbf{1996}, \emph{77}, 3865--3868\relax
\mciteBstWouldAddEndPuncttrue
\mciteSetBstMidEndSepPunct{\mcitedefaultmidpunct}
{\mcitedefaultendpunct}{\mcitedefaultseppunct}\relax
\EndOfBibitem
\bibitem[Grimme(2006)]{grimme-2006}
Grimme,~S. Semiempirical GGA-type density functional constructed with a
  long-range dispersion correction. \emph{J. Comput. Chem.} \textbf{2006},
  \emph{27}, 1787--1799\relax
\mciteBstWouldAddEndPuncttrue
\mciteSetBstMidEndSepPunct{\mcitedefaultmidpunct}
{\mcitedefaultendpunct}{\mcitedefaultseppunct}\relax
\EndOfBibitem
\bibitem[Goedecker \latin{et~al.}(1996)Goedecker, Teter, and
  Hutter]{goedcker_prb_1996}
Goedecker,~S.; Teter,~M.; Hutter,~J. Separable dual-space Gaussian
  pseudopotentials. \emph{Phys. Rev. B} \textbf{1996}, \emph{54},
  1703--1710\relax
\mciteBstWouldAddEndPuncttrue
\mciteSetBstMidEndSepPunct{\mcitedefaultmidpunct}
{\mcitedefaultendpunct}{\mcitedefaultseppunct}\relax
\EndOfBibitem
\bibitem[Hartwigsen \latin{et~al.}(1998)Hartwigsen, Goedecker, and
  Hutter]{goedcker_prb_1998}
Hartwigsen,~C.; Goedecker,~S.; Hutter,~J. Relativistic separable dual-space
  Gaussian pseudopotentials from H to Rn. \emph{Phys. Rev. B} \textbf{1998},
  \emph{58}, 3641--3662\relax
\mciteBstWouldAddEndPuncttrue
\mciteSetBstMidEndSepPunct{\mcitedefaultmidpunct}
{\mcitedefaultendpunct}{\mcitedefaultseppunct}\relax
\EndOfBibitem
\bibitem[Cholewinski \latin{et~al.}(2018)Cholewinski, Dixit, and
  Mpourmpakis]{chole-2018}
Cholewinski,~M.~C.; Dixit,~M.; Mpourmpakis,~G. Computational Study of Methane
  Activation on $\gamma$-Al$_2$O$_3$. \emph{ACS Omega} \textbf{2018}, \emph{3},
  18242--18250\relax
\mciteBstWouldAddEndPuncttrue
\mciteSetBstMidEndSepPunct{\mcitedefaultmidpunct}
{\mcitedefaultendpunct}{\mcitedefaultseppunct}\relax
\EndOfBibitem
\bibitem[Dixit \latin{et~al.}(2018)Dixit, Kostetskyy, and
  Mpourmpakis]{dixit-2018}
Dixit,~M.; Kostetskyy,~P.; Mpourmpakis,~G. Structure–Activity Relationships
  in Alkane Dehydrogenation on $\gamma$-Al$_2$O$_3$: Site-Dependent Reactions.
  \emph{ACS Catal.} \textbf{2018}, \emph{8}, 11570--11578\relax
\mciteBstWouldAddEndPuncttrue
\mciteSetBstMidEndSepPunct{\mcitedefaultmidpunct}
{\mcitedefaultendpunct}{\mcitedefaultseppunct}\relax
\EndOfBibitem
\bibitem[VandeVondele and Hutter(2003)VandeVondele, and Hutter]{joost_jcp_2003}
VandeVondele,~J.; Hutter,~J. An efficient orbital transformation method for
  electronic structure calculations. \emph{J. Chem. Phys.} \textbf{2003},
  \emph{118}, 4365--4369\relax
\mciteBstWouldAddEndPuncttrue
\mciteSetBstMidEndSepPunct{\mcitedefaultmidpunct}
{\mcitedefaultendpunct}{\mcitedefaultseppunct}\relax
\EndOfBibitem
\bibitem[Kolafa(2004)]{kolafa_jcc_2004}
Kolafa,~J. Time-reversible always stable predictor–corrector method for
  molecular dynamics of polarizable molecules. \emph{J. Comput. Chem.}
  \textbf{2004}, \emph{25}, 335--342\relax
\mciteBstWouldAddEndPuncttrue
\mciteSetBstMidEndSepPunct{\mcitedefaultmidpunct}
{\mcitedefaultendpunct}{\mcitedefaultseppunct}\relax
\EndOfBibitem
\bibitem[K\"uhne \latin{et~al.}(2007)K\"uhne, Krack, Mohamed, and
  Parrinello]{kuhne_prl_2007}
K\"uhne,~T.~D.; Krack,~M.; Mohamed,~F.~R.; Parrinello,~M. Efficient and
  Accurate Car-Parrinello-like Approach to Born-Oppenheimer Molecular Dynamics.
  \emph{Phys. Rev. Lett.} \textbf{2007}, \emph{98}, 066401\relax
\mciteBstWouldAddEndPuncttrue
\mciteSetBstMidEndSepPunct{\mcitedefaultmidpunct}
{\mcitedefaultendpunct}{\mcitedefaultseppunct}\relax
\EndOfBibitem
\bibitem[Nosé(1984)]{Nose_jcp_1994}
Nosé,~S. A unified formulation of the constant temperature molecular dynamics
  methods. \emph{J. Chem. Phys.} \textbf{1984}, \emph{81}, 511--519\relax
\mciteBstWouldAddEndPuncttrue
\mciteSetBstMidEndSepPunct{\mcitedefaultmidpunct}
{\mcitedefaultendpunct}{\mcitedefaultseppunct}\relax
\EndOfBibitem
\bibitem[Hoover(1985)]{Hoover_pra_1985}
Hoover,~W.~G. Canonical dynamics: Equilibrium phase-space distributions.
  \emph{Phys. Rev. A} \textbf{1985}, \emph{31}, 1695--1697\relax
\mciteBstWouldAddEndPuncttrue
\mciteSetBstMidEndSepPunct{\mcitedefaultmidpunct}
{\mcitedefaultendpunct}{\mcitedefaultseppunct}\relax
\EndOfBibitem
\bibitem[Digne \latin{et~al.}(2002)Digne, Sautet, Raybaud, Euzen, and
  Toulhoat]{dinge-2002}
Digne,~M.; Sautet,~P.; Raybaud,~P.; Euzen,~P.; Toulhoat,~H. Hydroxyl Groups on
  $\gamma$-Alumina Surfaces: A DFT Study. \emph{J. Catal.} \textbf{2002},
  \emph{211}, 1--5\relax
\mciteBstWouldAddEndPuncttrue
\mciteSetBstMidEndSepPunct{\mcitedefaultmidpunct}
{\mcitedefaultendpunct}{\mcitedefaultseppunct}\relax
\EndOfBibitem
\bibitem[Wilson(1979)]{wilson1979}
Wilson,~S. The dehydration of boehmite, $\gamma$-AlOOH, to
  $\gamma$-Al$_2$O$_3$. \emph{J. Solid State Chem.} \textbf{1979}, \emph{30},
  247--255\relax
\mciteBstWouldAddEndPuncttrue
\mciteSetBstMidEndSepPunct{\mcitedefaultmidpunct}
{\mcitedefaultendpunct}{\mcitedefaultseppunct}\relax
\EndOfBibitem
\bibitem[Biswas \latin{et~al.}(2020)Biswas, Kwon, Barsanti, Myllys, Smith, and
  Wong]{D0CP03832F}
Biswas,~S.; Kwon,~H.; Barsanti,~K.~C.; Myllys,~N.; Smith,~J.~N.; Wong,~B.~M. Ab
  initio metadynamics calculations of dimethylamine for probing pKb variations
  in bulk vs. surface environments. \emph{Phys. Chem. Chem. Phys.}
  \textbf{2020}, \emph{22}, 26265--26277\relax
\mciteBstWouldAddEndPuncttrue
\mciteSetBstMidEndSepPunct{\mcitedefaultmidpunct}
{\mcitedefaultendpunct}{\mcitedefaultseppunct}\relax
\EndOfBibitem
\bibitem[Barducci \latin{et~al.}(2008)Barducci, Bussi, and
  Parrinello]{barducci-2008}
Barducci,~A.; Bussi,~G.; Parrinello,~M. Well-Tempered Metadynamics: A Smoothly
  Converging and Tunable Free-Energy Method. \emph{Phys. Rev. Lett.}
  \textbf{2008}, \emph{100}, 020603\relax
\mciteBstWouldAddEndPuncttrue
\mciteSetBstMidEndSepPunct{\mcitedefaultmidpunct}
{\mcitedefaultendpunct}{\mcitedefaultseppunct}\relax
\EndOfBibitem
\bibitem[Biswas and Wong(2021)Biswas, and Wong]{BISWAS2021115624}
Biswas,~S.; Wong,~B.~M. Ab initio metadynamics calculations reveal complex
  interfacial effects in acetic acid deprotonation dynamics. \emph{J. Mol.
  Liq.} \textbf{2021}, \emph{330}, 115624\relax
\mciteBstWouldAddEndPuncttrue
\mciteSetBstMidEndSepPunct{\mcitedefaultmidpunct}
{\mcitedefaultendpunct}{\mcitedefaultseppunct}\relax
\EndOfBibitem
\bibitem[Mitchell \latin{et~al.}(1997)Mitchell, Sheinker, and
  Mintz]{mitchell-1997}
Mitchell,~M.~B.; Sheinker,~V.~N.; Mintz,~E.~A. Adsorption and Decomposition of
  Dimethyl Methylphosphonate on Metal Oxides. \emph{J. Phys. Chem. B}
  \textbf{1997}, \emph{101}, 11192--11203\relax
\mciteBstWouldAddEndPuncttrue
\mciteSetBstMidEndSepPunct{\mcitedefaultmidpunct}
{\mcitedefaultendpunct}{\mcitedefaultseppunct}\relax
\EndOfBibitem
\bibitem[Tsyshevsky \latin{et~al.}(2019)Tsyshevsky, Holdren, Eichhorn,
  Zachariah, and Kuklja]{roman-2019}
Tsyshevsky,~R.; Holdren,~S.; Eichhorn,~B.~W.; Zachariah,~M.~R.; Kuklja,~M.~M.
  Sarin Decomposition on Pristine and Hydroxylated ZnO: Quantum-Chemical
  Modeling. \emph{J. Phys. Chem. C} \textbf{2019}, \emph{123},
  26432--26441\relax
\mciteBstWouldAddEndPuncttrue
\mciteSetBstMidEndSepPunct{\mcitedefaultmidpunct}
{\mcitedefaultendpunct}{\mcitedefaultseppunct}\relax
\EndOfBibitem
\bibitem[Tsyshevsky \latin{et~al.}(2020)Tsyshevsky, Head, Trotochaud, Bluhm,
  and Kuklja]{roman-2020}
Tsyshevsky,~R.; Head,~A.~R.; Trotochaud,~L.; Bluhm,~H.; Kuklja,~M.~M.
  Mechanisms of Degradation of Toxic Nerve Agents: Quantum-chemical Insight
  into Interactions of Sarin and Soman with Molybdenum Dioxide. \emph{Surf.
  Sci.} \textbf{2020}, \emph{700}, 121639\relax
\mciteBstWouldAddEndPuncttrue
\mciteSetBstMidEndSepPunct{\mcitedefaultmidpunct}
{\mcitedefaultendpunct}{\mcitedefaultseppunct}\relax
\EndOfBibitem
\bibitem[Tsyshevsky \latin{et~al.}(2021)Tsyshevsky, McEntee, Durke, Karwacki,
  and Kuklja]{roman-2021}
Tsyshevsky,~R.; McEntee,~M.; Durke,~E.~M.; Karwacki,~C.; Kuklja,~M.~M.
  Degradation of Fatal Toxic Nerve Agents on Dry TiO$_2$. \emph{ACS Appl.
  Mater. Interfaces} \textbf{2021}, \emph{13}, 696--705\relax
\mciteBstWouldAddEndPuncttrue
\mciteSetBstMidEndSepPunct{\mcitedefaultmidpunct}
{\mcitedefaultendpunct}{\mcitedefaultseppunct}\relax
\EndOfBibitem
\bibitem[Wischert \latin{et~al.}(2011)Wischert, Copéret, Delbecq, and
  Sautet]{wischert-2011}
Wischert,~R.; Copéret,~C.; Delbecq,~F.; Sautet,~P. Dinitrogen: a selective
  probe for tri-coordinate Al “defect” sites on alumina. \emph{Chem.
  Commun.} \textbf{2011}, \emph{47}, 4890--4892\relax
\mciteBstWouldAddEndPuncttrue
\mciteSetBstMidEndSepPunct{\mcitedefaultmidpunct}
{\mcitedefaultendpunct}{\mcitedefaultseppunct}\relax
\EndOfBibitem
\bibitem[Zegers and Fisher(1998)Zegers, and Fisher]{zegers}
Zegers,~E.; Fisher,~E. Gas-Phase Pyrolysis of Diisopropyl Methylphosphonate.
  \emph{Combust. Flame} \textbf{1998}, \emph{115}, 230--240\relax
\mciteBstWouldAddEndPuncttrue
\mciteSetBstMidEndSepPunct{\mcitedefaultmidpunct}
{\mcitedefaultendpunct}{\mcitedefaultseppunct}\relax
\EndOfBibitem
\bibitem[Senyurt \latin{et~al.}(2020)Senyurt, Schoenitz, and
  Dreizin]{senyurt-2020}
Senyurt,~E.~I.; Schoenitz,~M.; Dreizin,~E.~L. Rapid destruction of sarin
  surrogates by gas phase reactions with focus on diisopropyl methylphosphonate
  (DIMP). \emph{Def. Technol.} \textbf{2020}, \relax
\mciteBstWouldAddEndPunctfalse
\mciteSetBstMidEndSepPunct{\mcitedefaultmidpunct}
{}{\mcitedefaultseppunct}\relax
\EndOfBibitem
\bibitem[Yuan and Eilers(2019)Yuan, and Eilers]{yuan-2019}
Yuan,~B.; Eilers,~H. T-jump pyrolysis and combustion of diisopropyl
  methylphosphonate. \emph{Combust. Flame} \textbf{2019}, \emph{199},
  69--84\relax
\mciteBstWouldAddEndPuncttrue
\mciteSetBstMidEndSepPunct{\mcitedefaultmidpunct}
{\mcitedefaultendpunct}{\mcitedefaultseppunct}\relax
\EndOfBibitem
\bibitem[Mukhopadhyay \latin{et~al.}(2021)Mukhopadhyay, Schoenitz, and
  Dreizin]{MUKHOPADHYAY2021}
Mukhopadhyay,~S.; Schoenitz,~M.; Dreizin,~E.~L. Vapor-phase decomposition of
  dimethyl methylphosphonate (DMMP), a sarin surrogate, in presence of metal
  oxides. \emph{Def. Technol.} \textbf{2021}, \emph{17}, 1095--1114\relax
\mciteBstWouldAddEndPuncttrue
\mciteSetBstMidEndSepPunct{\mcitedefaultmidpunct}
{\mcitedefaultendpunct}{\mcitedefaultseppunct}\relax
\EndOfBibitem
\bibitem[Neupane \latin{et~al.}(2018)Neupane, Barnes, Barak, Ninnemann, Loparo,
  Masunov, and Vasu]{neupane-2018}
Neupane,~S.; Barnes,~F.; Barak,~S.; Ninnemann,~E.; Loparo,~Z.; Masunov,~A.~E.;
  Vasu,~S.~S. Shock Tube/Laser Absorption and Kinetic Modeling Study of
  Triethyl Phosphate Combustion. \emph{J. Phys. Chem. A} \textbf{2018},
  \emph{122}, 3829--3836\relax
\mciteBstWouldAddEndPuncttrue
\mciteSetBstMidEndSepPunct{\mcitedefaultmidpunct}
{\mcitedefaultendpunct}{\mcitedefaultseppunct}\relax
\EndOfBibitem
\bibitem[Mathieu \latin{et~al.}(2019)Mathieu, Kulatilaka, and
  Petersen]{mathieu-2019}
Mathieu,~O.; Kulatilaka,~W.~D.; Petersen,~E.~L. Shock-tube studies of Sarin
  surrogates. \emph{Shock Waves} \textbf{2019}, \emph{29}, 441--449\relax
\mciteBstWouldAddEndPuncttrue
\mciteSetBstMidEndSepPunct{\mcitedefaultmidpunct}
{\mcitedefaultendpunct}{\mcitedefaultseppunct}\relax
\EndOfBibitem
\end{mcitethebibliography}

\section*{TOC graphic}
\begin{center}
    \includegraphics[width=15cm]{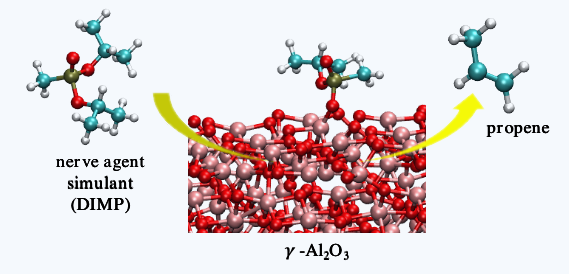}
\end{center}

\end{document}